\documentclass[11pt,a4paper]{article}
\usepackage{jheppub}

\usepackage{graphicx}
\usepackage{amsmath,amssymb,amsthm}
\usepackage{latexsym,graphicx,color,subfigure}
\usepackage{bm}

\newcommand{\be}{\begin{equation}}
\newcommand{\ee}{\end{equation}} 
\newcommand{\beq}{\begin{eqnarray}}
\newcommand{\eeq}{\end{eqnarray}}

\newcommand{\D}{\mathcal{D}}
\newcommand{\p}{\partial}
\newcommand{\Tr}{{\rm Tr}}

\newcommand{\bea}{\begin{eqnarray}}
\newcommand{\eea}{\end{eqnarray}}
\def\Tr{ \hbox{\rm Tr}}

\def\half{\frac{1}{2}}
\def\a{\alpha}

\def\rh{\rho}
\def\half{\frac{1}{2}}

\def\l{\left}
\def\r{\right}
\def\Dg{\Delta_g}

\title{{\bf  The  effective action of a BPS Alice string}}
\author{Chandrasekhar Chatterjee$^{1, a}$} 
\author{Muneto Nitta$^{1,b}$}
 \affiliation[a]{
 Department of Physics, and Research and Education Center for Natural Sciences,\\ Keio University,Hiyoshi 4-1-1, Yokohama, Kanagawa 223-8521, Japan}
 \emailAdd{chandra.chttrj@gmail.com$^{a}$, chandra@phys-h.keio.ac.jp$^{a}$}
 \emailAdd{nitta(at)phys-h.keio.ac.jp$^{b}$}

\date{\today}

\abstract{
Recently a BPS Alice string has been found in  a $U(1)\times SU(2)$ gauge theory coupled with a charged complex adjoint scalar field 
\cite{Chatterjee:2017jsi}.
It is a half BPS state preserving a half of 
supercharges when embedded into a supersymmetric gauge theory. 
In this paper, we study zero modes of a BPS Alice string.
After presenting $U(1)$ and translational zero modes, 
we construct the effective action of these modes.
In contrast to previous analysis of the conventional Alice string 
for which only large distance behaviors are known, 
we can perform calculation exactly in the full space 
thanks to BPS properties.
}

\begin{document}
\maketitle

\section{Introduction}
Topological vortices are an important and interesting subject to study not only because of their mathematical elegance but also of practical purposes 
from superfluids, superconductors, ultracold atomic gases 
to quantum field theory, QCD, string theory and cosmology. 
In cosmological context, a symmetry breaking phase transition may have occurred in the early universe due to rapid cooling and expansion of universe, during which topological vortices known as `cosmic strings' 
may be created \cite{Kibble:1976sj, Hindmarsh:1994re}.  
Such classical vortex configurations become more interesting when interacting with quantum fields and these interactions generate massless excitations (zero modes) near the vortex core and dictate the low-energy dynamics of vortices. 
When a $U(1)$ zero mode arises due to a breaking of bulk continuous symmetry inside the vortex core, the string core behaves as a superconducting wire \cite{Witten:1984eb}. In the case of local symmetry the excitation of zero mode inside the vortex may also excite massless gauge field in the bulk which generates logarithmically divergent energy. 

Among those, Alice strings have very peculiar features. 
When a charged particle encircles an Alice string, 
its electric charge changes the sign 
\cite{Schwarz:1982ec}, 
and therefore one cannot define an electric charge globally.  
The simplest example of an Alice string can be found in an 
$SO(3)$ gauge theory with 5 representation,
in which the $SO(3)$ gauge 
group is spontaneously broken down to $O(2)$ 
in the vacuum.
So the vacuum manifold or order parameter space becomes
$G/H =SO(3)/O(2) \simeq {\mathbb R}P^2$. 
This allows a non-trivial homotopy group 
$\pi_1({\mathbb R}P^2) \simeq {\mathbb Z}_2$ 
supporting an Alice string.
The unbroken generator can be identified as the electromagnetic $U(1)$ generator, 
and its sign changes while it encircles 
the Alice string once, as mentioned above.
The bulk system of the Alice string behaves as `Alice electrodynamics' where charge conjugation ($\mathbb{Z}_2$) is taken as a local symmetry \cite{Kiskis:1978ed}. 
A change of the sign of electric charges around the Alice string 
implies a lost of the charge. 
This lost can be explained in terms of a nonlocal charge called 
the ``Cheshire charge'' 
\cite{Preskill:1990bm, Bucher:1993jj, Bucher:1992bd}.\footnote{
The Alice string is also known to  exchange magnetic charges from magnetic particles which creates the magnetic Cheshire  charges \cite{Bucher:1992bd}. Related to this,
a ring of the Alice string can be thought of  as magnetic monopole at large distances when the $U(1)$ modulus is twisted along the ring, 
and this is supported by the   
 non-trivial second homotopy group, $\pi_2(G/H)\ne 0$ \cite{Shankar:1976un,  Benson:2004ue, Bais:2002ae, Striet:2003na}.  
}
Many interesting global phenomena arise due to this exotic property, like the generalized Aharonov-Bohm effect \cite{Alford:1988sj}, anyonic exchange statistics  etc \cite{Lo:1993hp}. 
 It was suggested that a nature of the Cheshire charge 
 can be explained as follows;
The $U(1)$ gauge symmetry in the bulk 
is spontaneously broken 
in the presence of an Alice string, thereby giving rise to 
a $U(1)$ zero mode, 
in addition to usual translational zero modes. 
This $U(1)$ zero mode is non-normalizable and is 
charged under the $U(1)$ gauge group.
If there are two or more Alice strings, the Cheshire charge 
should be nonlocally defined over them in terms of 
a zero mode attached to each string. 
However, the unbroken bulk symmetries are multivalued
in the case of Alice strings, and therefore 
the behavior of zero mode analysis becomes little subtle, in which case one needs to carefully take into account the multivalued nature of the unbroken group generator in the analysis, which is called ``obstruction'' \cite{Alford:1990mk, Alford:1990ur, Alford:1992yx, Bolognesi:2015ida}.\footnote{Topological obstruction for monopole is  discussed in Ref.~\cite{Nelson:1983bu,Abouelsaood:1982dz,Balachandran:1982gt}.} 
In general for vortices, this occurs due to the existence of 
{\it local} discrete  unbroken symmetries \cite{Krauss:1988zc}.

One of difficulty to perform concrete calculation of zero mode analysis 
in a multi-string system may be due to the existence 
of the interaction between Alice strings.
The homotopy group
$\pi_1({\mathbb R}P^2) \simeq {\mathbb Z}_2$ implying that 
two Alice strings can annihilate each other  
and the existence of an
attraction between them.
One usually manipulate for instance an exchange of strings 
as adiabatic process but it is dynamically difficult or impossible.
In contrast, there is no force between 
Bogomolnyi-Prasad-Sommerfield (BPS) solitons (strings)  
 \cite{Bogomolny:1975de,Prasad:1975kr}.
Therefore, one can place them at arbitrary positions, 
and $n$ strings have the total energy (tension) exactly $n$ times larger than 
that of a single string, 
thereby allowing a large moduli space of multi-string configurations.
However, 
in contrast to the conventional strings
in the Abelian-Higgs model 
\cite{Abrikosov:1956sx,Nielsen:1973cs}, which can become 
BPS at the critical coupling,  
previously studied Alice strings were all 
non-BPS because  
two Alice strings annihilate each other 
as mentioned above.
In our recent work \cite{Chatterjee:2017jsi}, 
we have found that an 
$SU(2) \times U(1)$ gauge theory coupled 
with one charged complex adjoint scalar field admits 
BPS Alice strings, 
and have shown that it preserves a half supersymmetry 
if embedded into supersymmetric gauge theory. 
In this theory, one can expect to place 
as many Alice strings as one likes 
at arbitrary positions, thereby openings 
a possibility to study exotic phenomena of Alice strings 
more concretely.

In the present paper, as a first step, 
we systematically study zero mode analysis of 
a single BPS Alice string. 
A zero mode analysis for a conventional Alice string has been 
already discussed in literature 
but only at large distance since equations cannot be solved exactly.  In this work we shall show that
the zero modes of a BPS Alice string can be solved exactly as 
functions of the vortex profile functions, so that behaviors 
of the zero modes can be understood on the full space. 
We then construct the effective action for the $U(1)$ and translational zero 
modes of a single BPS Alice string.

This paper is organized as follows.
  In Sec.~\ref{review}, we give a brief review of BPS Alice strings. 
  In Sec.~\ref{modes}, we discuss the translational and $U(1)$ moduli  separately. In Sec.~\ref{effectiveaction}  we derive the effective action 
  of zero modes, and finally we give a summary of our results 
  and discussions in Sec.~\ref{discussion}, in which 
  we discuss a possible interaction of 
 the $U(1)$ mode and the $U(1)$ gauge field in the bulk.

\section{BPS Alice strings}\label{review}
In this section, we give a short review of BPS Alice strings, 
for details see Ref.~\cite{Chatterjee:2017jsi}. 
We consider an $SU(2)\times U(1)_b$ gauge theory 
with $SU(2)$ and $U(1)$ gauge fields $A_{\mu}$ and $a_{\mu}$, 
respectively, coupled  
with charged complex scalar fields $\Phi$ in the adjoint representation.
The corresponding action can be written as
\begin{eqnarray}
 \label{action}
&&{I} = \int d^4x \left[- \frac{1}{2} \Tr F_{\mu\nu} F^{\mu\nu}- \frac{1}{4} f_{\mu\nu}f^{\mu\nu} + \Tr | D_\mu\Phi|^2  
-  \frac{\lambda_g}{4} \Tr[\Phi,\Phi^\dagger]^2 
-  \frac{\lambda_e}{2}\left(\Tr \Phi\Phi^\dagger  - 2 \xi^2\right)^2\right]. \nonumber \\ \end{eqnarray}
where
$D_\mu\Phi = \p_\mu\Phi - i  e a_\mu\Phi -i g \left[A_\mu, \Phi\right]$, 
and 
$F_{\mu\nu} = \p_\mu A_\nu - \p_\nu A_\mu - i g [A_\mu, A_\nu], \, f_{\mu\nu} =  \p_\mu a_\nu - \p_\nu a_\mu$,  
with  gauge couplings  $g$ and $e$ for 
$SU(2)$ and  $U(1)_b$ gauge fields, respectively.\footnote{
The cosmic strings in the same theory were also discussed before in Ref.~\cite{Davis:1997ny}. However explicit Bogomol'nyi completion and BPS vortex solutions were not  discussed there, and they seem to be 
unaware of the fact that they are 
Alice strings even though the same theory. 
}
The $U(1)_b$ symmetry gives the stability to a fractional vortex and we can have an unbroken $\mathbb{Z}_2$ in the vacuum. We can choose the vacuum expectation value of the field $\Phi$ as 
\begin{eqnarray}\label{phivac}
\langle\Phi\rangle_v = 2 \xi \tau^1,\;
\end{eqnarray}
with $\tau^1 = \half \sigma^1$. This breaks the gauge symmetry group $G = U(1)_b \times \frac{SU(2)}{\mathbb{Z}_2} \simeq U(1)_b \times SO(3)$ as
\begin{eqnarray}
 G = U(1)_b \times \frac{SU(2)}{\mathbb{Z}_2}  \simeq U(1)_b \times SO(3) \longrightarrow  
 H = \mathbb{Z}_2 \ltimes  U(1)_1 \simeq O(2),
\end{eqnarray}
where $\ltimes$ stands for a semi-direct product. This can be understood as follows. 
Eq.~(\ref{phivac}) says that any rotation around 
$\tau^1$ keeps $\langle\Phi\rangle_v$ invariant. The  unbroken discrete group $ \mathbb{Z}_2$ is defined as
a simultaneous $\pi$ rotation around axes directed along any  linear combination of $\tau^3$ and $\tau^2$ and in $U(1)_b$. This keeps $\langle\Phi\rangle_v$ invariant since both the $\pi$ rotations generate sign changes separately. The unbroken group elements can be written as
\begin{eqnarray}
H = \l\{\left(1, \,\, e^{\frac{i\alpha}{2}\sigma^1}\right),\, \left(-1, \,\,i\left(c_2 \sigma^2 + c_3\sigma^3\right)e^{i\frac{\alpha}{2}\sigma^1}\right)\r\},
\end{eqnarray}
where $c_2, c_3$ are arbitrary real constants normalized to the unity as $c_2^2 + c_3^2 =1$.
The semi-direct product is arising here because 
${\mathbb Z}_2$ element changes the action of $U(1)_1$.
The fundamental group for this symmetry breaking 
can be written as
\begin{eqnarray}
 \pi_1  \left( \frac{U(1)_b \times SO(3)}{O(2)}\right)
\simeq 
 \pi_1  \left( \frac{S^1 \times S^2}{\mathbb{Z}_2}\right)
\simeq \mathbb{Z}.
\end{eqnarray}
This nontrivial fundamental group supports the existence of stable strings. The $\mathbb{Z}_2$ element makes the generator of unbroken $U(1)_1$ globally undefined in the presence of a string:
the generator of the unbroken $U(1)$ changes sign as it encircles 
the string once, and this  property of the vortex in this system identifies it as an Alice string. 

Here we are studying BPS vortices  and 
the BPS completion can be performed if we consider the critical couplings 
$ \lambda_e = e^2$ and $\lambda_g = g^2$.  
In this case, the theory can be embedded into a supersymmetric gauge 
theory in which Alice strings preserve a half of the supersymmetry \cite{Chatterjee:2017jsi}, but we do not focus on such the aspect in this paper. 
The Bogomol'nyi completion of the tension, that is the static energy per a unit length, is found to be
\begin{eqnarray}
\label{tension}
\mathcal{T} & =& \int d^2x \left[\Tr\left[ F_{12} \pm \frac{g}{2}[\Phi,\Phi^\dagger]\right]^2 + \Tr|\D_{\pm} \Phi|^2 + \half\left[  f_{12} \pm e \left(\Tr \Phi\Phi^\dagger - 2 \xi^2\right)\right]^2
\pm 2 e f_{12} \xi^2 \right.\Big{]} \nonumber\\ 
 &\ge& 2 e  \xi^2 \left|\int d^2x\; f_{12}\right|,\;
\end{eqnarray}
with $\D_{\pm} \equiv \frac{D_1 \pm i D_2}{2}$. The saturation of the inequality  in the above equation, 
implying the minimum tension of vortex solution
in the same topological sector, yield BPS
equations given by
\begin{eqnarray}
&&
 f_{12} \pm e \left(\Tr \Phi\Phi^\dagger - 2 \xi^2\right) =0,
 \label{eq:BPS1}
 \;\\
&&F_{12} \pm \frac{g}{2}[\Phi,\Phi^\dagger] =0,\;
\label{eq:BPS2}
\\
&& \D_{\pm} \Phi  = \left(\D_{\pm} \Phi\right)^\dagger   = 0.\label{eq:BPS3}
\end{eqnarray}

In order to solve these equations for a single vortex solution, we take the vortex ansatz  of the scalar and gauge fields as
\begin{eqnarray}\label{phivortex}
&&\Phi(r, \theta) = 
\xi\left(
\begin{array}{ccc}
 0 & f_1(r) e^{i\theta}     \\
f_2(r)   & 0  \end{array}
\right),\\
&&a_i (r, \theta) = - \frac{1}{2e} \frac{\epsilon_{ijx_j}}{r^2}a(r), \quad
A_i(r, \varphi) = -\frac{1}{4g} \frac{\epsilon_{ijx_j}}{r^2}\sigma^3A(r),
\end{eqnarray}
where $\{r, \theta \}$ are radial and angular coordinates 
of the two dimensional space, 
respectively. 
The equations for the profile functions
$f_1(r), f_2(r), A(r)$ and $a(r)$ depending only on the radial coordinate 
are the same with those of a non-Abelian vortices 
\cite{Hanany:2003hp,Auzzi:2003fs,Eto:2005yh,Eto:2006pg, Tong:2008qd, Shifman:2007ce},
and they 
can be solved numerically 
with the boundary conditions
\begin{eqnarray}
f_1(0) = f_2'(0) = 0,&&\quad f_1(\infty) = f_2(\infty) =1,\\ A(0)= a(0) = 0,&&\quad A(\infty)= a(\infty) = 1.
\end{eqnarray}
The numerical solution is displayed in the Fig.~\ref{profile}.
\begin{figure}[!htb]
\label{profile}
\centering
\subfigure{\includegraphics[totalheight=3.5cm]{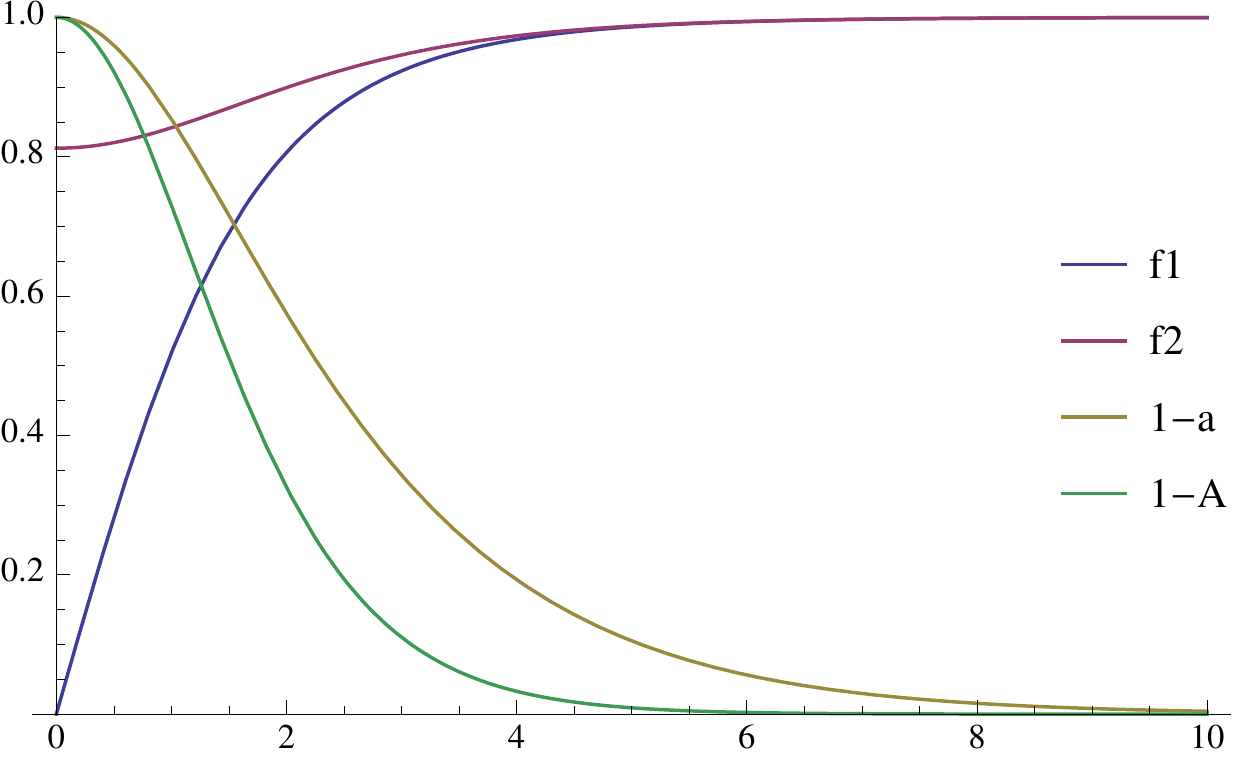}}
\caption{The numerical solutions of the profile functions $f_1(r)$ ,  $f_2(r)$, $1 -a(r)$ and $1-A(r)$ are displayed  for a vortex configuration of winding number one as a function of $r$(the distance  from the vortex center) for  $ l = \frac{e}{g}= 0.5 $  \cite{Chatterjee:2017jsi} . }
\label{profile}
\end{figure}

\subsection*{Unbroken $U(1)$ symmetry}
The vacuum configuration in Eq.~(\ref{phivac}) shows that it is invariant under the adjoint action of the $U(1)$ subgroup generated by $\tau^1 = \half \sigma^1$. However, in the presence of a string, the scalar field configuration in Eq.~(\ref{phivortex}) depends on space coordinates even at very large distances as 
\begin{align}
\Phi(R, \varphi) & \sim \xi e^{i\frac{\theta}{2}} \left(
\begin{array}{ccc}
 0 &  e^{i\frac{\theta}{2}}     \\
  e^{-i\frac{\theta}{2}} & 0  \end{array}
\right) 
=  \Omega_0(\theta) \Omega_3(\theta) \Phi(R, 0) \Omega_3^{-1}(\theta), 
\end{align}
with $\Phi(R, \theta=0) = \xi \sigma^1$. 
The holonomy $\Omega$ can be defined by
\begin{eqnarray}
\label{holonomy}
\Omega_0(\theta) = e^{ie \int_0^\theta {\bf a\cdot dl}} = e^{i\frac{\theta}{2}}, \quad \Omega_3(\theta) = Pe^{i g \int_0^\theta {\bf A\cdot dl}} = e^{i\frac{\theta}{4}\sigma^3} .
\end{eqnarray}
So the embedding of the unbroken $U(1)$ group becomes space dependent and the generator must be changed by the holonomies as it goes around the string as 
\begin{eqnarray}
Q_\theta = \Omega_3(\theta) Q_0 \Omega_3(\theta)^{-1}.
\end{eqnarray}
After a full encirclement it is easy to find that $\Omega_3(2\pi)\in \mathbb{Z}_2$ and
\begin{eqnarray}
Q_{2\pi} =  - Q_0 = e^{i2\pi\zeta } Q_0,
\end{eqnarray}
where the `obstruction' parameter $\zeta = \half$. 
We shall see that the $\mathbb{Z}_2$ affects the zero mode solution in next section.

One comment is in order.
A global version of the Alice string in our theory 
was already found in Ref.~\cite{Leonhardt:2000km} 
(see Refs.~\cite{Kobayashi:2011xb,Kawaguchi:2012ii}) 
in the context of Bose-Einstein condensation 
of ultra cold atomic gases. 
In the same context, a global monopole as 
a twisted Alice ring was also found 
\cite{Ruostekoski:2003qx}.
By gauging $U(1) \times SU(2)$ symmetry, we obtained our theory 
which allows BPS Alice strings for the critical coupling 
\cite{Chatterjee:2017jsi}.

\section{The moduli of a BPS Alice string}\label{modes}
Zero modes arise due to the breaking of any continuous unbroken bulk symmetry in the presence of a vortex solution. We can have continuously 
degenerate BPS solutions which keeps BPS equation and tension invariant. So we have moduli space of solutions, 
and the motion on the moduli space generates zero modes. 
Here we would like to discuss translational and $U(1)$ zero modes separately.

\subsection{Translational moduli}
 In this subsection, we discuss translational modes 
 arising due to breaking of translational invariance,
 which are are almost the same as other BPS vortices. 
 The solution of the BPS equations can be shifted to any arbitrary point as center of vortex. So the zero modes can be found by expanding the BPS solutions around the origin which is taken as center of our vortex. In a particular gauge the  zero mode solutions look like
 \begin{eqnarray}
a_i^T (x) = - f_{mi} \delta X_m, \quad A_i^T(x) = - F_{mi}(x) \delta X_m, \quad \Phi^T (x) = - \D_m \Phi(x) \delta X_m,
\end{eqnarray}
where $f_{mi}$ and $F_{mi}(x)$ are the field strengths computed from  the BPS solutions and $ \delta X_m$ is the displacement of the centre of the vortex from the origin, where we place the vortex. 

\subsection{$U(1)$ modulus}
As it is well known  that zero modes arise when a soliton solution is not invariant under the action of a continuous unbroken symmetry group. In the case of the Alice string the unbroken $U(1)$ symmetry in the bulk 
is spontaneously broken in the core of the vortex.  It can be observed easily when applying 
$U(1)$ transformation on  the order parameter. On the $x$-axis (at $\theta =0$) the order parameter can be written as
\begin{eqnarray}
\Phi(r, 0) = 
\xi\left(
\begin{array}{ccc}
 0 & f_1(r)     \\
f_2(r)   & 0  \end{array}
\right).
\end{eqnarray}
Any small change due to the  action of the $U(1)$ group elements $e^{i\frac{\varphi}{2} \sigma^1 }$ for small $\varphi$ is written as
\begin{align}
\delta\Phi(r, 0) =  i\frac{\varphi}{2} \Big[\sigma^1, \Phi(r,0)\Big] = i\frac{\varphi\xi}{2} \Big(f_2(r) - f_1(r)\Big)\sigma^3 . 
\end{align}
At the large distances from the vortex center, the order parameter is invariant under the $U(1)$ action because of $f_1(\infty) = f_2(\infty) =1$.   It is, however, not the case around the vortex core because of 
$f_1(r) \ne f_2(r)$ inside the vortex, according the solution displayed in 
Fig.~\ref{profile}.  The $U(1)$  transformation changes the magnetic flux as well at the vortex core, and so it generates a physically distinct degenerate solution. Namely, if $\{\Phi, {A}_i\}$  minimizes the  energy, then 
$\{\Phi(\varphi), {A_i(\varphi})\}$ also does, where 
$\{\Phi(\varphi), {A_i(\varphi})\}$ is defined by a global $U(1)$ transformation,
\begin{eqnarray}
\Phi(\varphi) =  U_\varphi \Phi U_\varphi^\dagger, \qquad{A_i(\varphi)} = U_\varphi A_i U^\dagger_\varphi, \qquad U_\varphi \in  U(1),
\end{eqnarray}
for a constant parameter $\varphi$. 
It is also true if $\varphi$ is a function of the $x, y$-coordinates, 
in which case we have to add an inhomogeneous term with the gauge field
 \cite{Alford:1990ur}. Since the $U(1)$ symmetry is broken inside the vortex we stick to a global transformation. This would generate an infinite set of solutions which are physically distinct. So $\varphi$ can be treated as a $U(1)$ modulus.

\section{The effective action of a BPS Alice string}\label{effectiveaction}
In the last section we  have discussed the existence of the zero modes by heuristic arguments. 
In this section, we construct the effective action of these zero modes by 
the moduli approximation \cite{Manton:1981mp}.

\subsection{The effective action of the translational moduli}
  Let us start by shifting the center of the vortex to 
  $(X(z,t), Y(z,t))$, and then the solution of BPS equations becomes  
\begin{eqnarray}
 a_i(x_m-X_m(z,t)) , \quad  
 A_i(x_m-X_m(z,t)) , \quad 
 \Phi(x_m - X_m(z,t)).
\end{eqnarray}
Here, we assume a slow variation of the vortex center with 
the moduli space approximation. This slow variation generates non-zero current along the $t$-$z$ directions, 
and we have to introduce $t, z$ derivatives to solve the Gauss' law.  
We may write the $\alpha = \{0, 3\}$ derivative terms as 
\begin{eqnarray}
f_{i\alpha} = V^j_\alpha \p_j a_i,  \quad 
F_{i\alpha} = V^j_\alpha\p_j A_i,\quad  
\p_\alpha\Phi =  - V^j_\alpha\p_j\Phi.
\end{eqnarray}
Here the velocity is defined as $V^j_\alpha = \p_\a X^j$. Let us choose a gauge so that above expressions can be rewritten as 
\begin{eqnarray}
f_{i\alpha} &=& V^j_\alpha f_{ji}, \quad F_{i\alpha} =  V^j_\alpha F_{ji}, \quad \p_\alpha\Phi =  - V^j_\alpha\D_j\Phi .
\end{eqnarray}
Here we are considering only small fluctuations, and we have neglected all higher order terms. Now using the first two terms, we may write
\begin{eqnarray}
\half f_{i\a}^2 = \frac{1}{4}\l(V^k_\a\r)^2 f_{ij}^2,\qquad \Tr F_{i\a}^2 = \half\l(V^k_\a\r)^2 \Tr F_{ij}^2.
\end{eqnarray}
By using the BPS equations, 
we may rewrite above equations as
\begin{eqnarray}\label{f2}
\half f_{i\a}^2 &=& \frac{1}{2}\l(V^k_\a\r)^2 \l[ \frac{1}{4} f_{ij}^2 + \frac{e^2}{2} \left(\Tr \Phi\Phi^\dagger - 2 \xi^2\right)^2 \r], \\ \label{F2}
\Tr F_{i\a}^2 &=& \half\l(V^k_\a\r)^2 \left[\half  \Tr F_{ij}^2 + \frac{g^2}{4}\Tr[\Phi,\Phi^\dagger]^2\r]. 
\end{eqnarray}
Now using rotational invariance and BPS equation $\D_+\Phi =0$ we may rewrite,
\begin{eqnarray}\label{p2}
\int d^2x |\p_\alpha\Phi|^2 = \half\l(V^k_\a\r)^2\int d^2x \Tr |\D_i\Phi|^2.
\end{eqnarray}
 We now use the equations (\ref{f2}) , (\ref{F2}) and (\ref{p2}) and following Eq.~(\ref{tension}) and the solutions of BPS equations the effective action is written as
 \begin{eqnarray}
&&\int d^2x\l[\half f_{i\a}^2 + \Tr F_{i\a}^2 + \Tr |\D_\alpha\Phi|^2 \r] 
= \pi \xi^2 \l(\p_\a X^k\r)^2.
\end{eqnarray}


\subsection{The effective action of the $U(1)$ modulus}

 To understand the behavior of the  $U(1)$ modulus we need to excite the modes along the $z$-axis to write down the effective action.  We set the fluctuation of the $U(1)$ modulus of the scalar field as 
\begin{eqnarray}
\Phi({\varphi}(t, z), r, \theta=0) &=& U_\varphi(t, z) \Phi(r)U^\dagger_\varphi(t,z), \qquad U_\varphi(t, z) = e^{i\varphi(t,z)\tau^1}, 
\end{eqnarray}
where $\varphi(t, z)$ is taken as a slowly varying function on the $t$-$z$ plane. Since the generator of unbroken group  is not well defined  globally as described in Ref.~\cite{Chatterjee:2017jsi}, we define the $\theta$ dependence of the scalar as
\begin{eqnarray}
 \Phi({\varphi}(t, z), r, \theta) &=&  e^{\frac{\theta}{2}}U_\varphi(\theta, t, z)\Phi({\varphi}(t, z), r, \theta=0) U_\varphi^\dagger(\theta, t, z), \quad U_\varphi(\theta, t, z) =  e^{\frac{i\theta}{4}U_\varphi \sigma^3 U^\dagger_\varphi}.\qquad\,
\end{eqnarray}
This can be rewritten as 
\begin{eqnarray}
 \Phi({\varphi}(t, z), r, \theta)  = e^{i\frac\varphi 2 \sigma^1 }\left(
\begin{array}{ccc}
 0 & f_1(r)e^{i\theta}    \\
f_2(r)   & 0  \end{array}
\right)e^{-i\frac\varphi 2 \sigma^1 } =  U_\varphi(t, z) \Phi(r, \theta) U_\varphi^\dagger(t, z).
\label{phitz}
\end{eqnarray}
Similarly we can introduce the gauge field  fluctuation corresponding to $U(1)$ modulus as
\begin{eqnarray}\label{Atz}
A_i (\varphi(t, z), x, y) = -\frac{1}{4g} \frac{\epsilon_{ijx_j}}{r^2}A(r) \,\,\,U_\varphi(t, z) \sigma^3U_\varphi^\dagger(t, z).
\end{eqnarray}
The $(t, z)$ dependence of the fluctuations may keep the BPS equations unchanged, but it would generate $J_0$ and $J_3$ in the equations of motion of the gauge field. 
This violates the Gauss' law and Biot-Savart-law, and we need to introduce $A_0$ and $A_3$ to solve these equations.  Since $F_{\alpha\beta} = 0$, we  
may write the effective action as 
\begin{eqnarray}\label{eff1}
I_{\rm eff} = \int dtdz\left[ \int d^2x \left\{-\half f_{i\alpha}f^{i\alpha} - \Tr F_{i\alpha}F^{i\alpha} + \Tr |D_\alpha \Phi(\varphi)|^2 \r\}\r], \quad \alpha = \{0, 3\},
\end{eqnarray}
where
$
F_{i\alpha} = \p_iA_\a - D_\a A_i(\varphi), D_\a = \p_\a -ig [A_\a,\, \cdot\,\,], D_\a \Phi(\varphi) = \p_\a \Phi(\varphi) - i e a_\a \Phi(\varphi) -ig [A_\a, \Phi(\varphi)]$.
Eq.~(\ref{eff1})  is a part of the full action, and the rests vanish for the BPS solutions.

Now by variations of $a_\a$ and $A_\a$, we may write the equations of motion as 
\begin{eqnarray}\label{Gausslaw}
\p_if_{i\a} &=& i e \Tr\l[\Phi^\dagger D_\a \Phi - \Phi \l(D_\a \Phi\r)^\dagger\r] ,\nonumber\\
D_iF_{i\a} & = & -i \frac{g}{2} \l\{  \l[\Phi^\dagger, D_\a \Phi\r] + \l[ \Phi, (D_\a \Phi)^\dagger\r] \r\}.
\end{eqnarray} 
To solve these equations of motion we introduce an ansatz of 
generated components $A_\a$ of the gauge field as
\begin{eqnarray}\label{ansatz}
A_\a(t, z, r, \theta) &=& \frac{1}{g}\l[\l(1 - \Psi_1(r, \theta)\r) \tau^1 + \Psi_2(r, \theta) T^2\r]\p_\a\varphi(t, z), \qquad  \\ 
a_\a &=& 0,\qquad T^2 = U_\varphi(t, z)\, \tau^2\, U_\varphi ^\dagger(t, z), \qquad \tau^i = \half \sigma^i.\nonumber
\end{eqnarray}
Here $\Psi_i(r, \theta)$ are real and imaginary components of 
an arbitrary complex function on the $x$-$y$ plane and can be found by solving equations of motion.
To find the profile functions $\Psi_i(r, \theta)$, we may insert the ansatz into the equations of motion in Eq.~(\ref{Gausslaw}) and solve 
the equations for $\Psi_i(r, \theta)$. 
Likewise we may insert 
the ansatz into the effective action in Eq.~(\ref{eff1}) and find equations for $ \Psi_i$ after a variation with respect to $\Psi_i$. Since both the cases give the same equations of $\Psi_i$,
we use the second calculation here. It is straightforward  to show that the above ansatz in  Eq.~(\ref{ansatz}) solves the equations of motion in Eq.~(\ref{Gausslaw}). 
Now we may use the zero mode ansatz in Eq.~(\ref{ansatz}) and  the vortex solution ansatz in Eq.~(\ref{phitz}) and (\ref{Atz}) to express $F_{i\a}$ and $D_\a\Phi$  in terms of 
$\Psi_i$ and $\varphi$ as
\begin{eqnarray}
F_{i\a}&&=\frac{\p_\a \varphi}{g}\l[  - \l( \p_i \Psi_1 + \zeta  b_i\Psi_2  \r)\tau^1 + \l( \p_i\Psi_2 - \zeta b_i  \Psi_1\r) T^2\r], \nonumber\\
D_\a \Phi &&= - e^{i\frac{\theta}{2}}g \xi\l[ \Psi_1 \,\Phi_2   +   \Psi_2 \, \Phi_1 \r]\p_\a\varphi T^3,\label{Fphitz}
\end{eqnarray}
where the ``obstruction" parameter is $\zeta= \half$ ,   $b_i = - \frac{\epsilon_{ij} x_j}{r^2} A(r)$ and $T^3 = U_\varphi(t, z)\, \tau^3\, U_\varphi ^\dagger(t, z)$. Here $\Phi_1$ and $\Phi_2$ have been defined by
\begin{eqnarray}
\Phi_1(r, \theta) = \l(f_1(r) e^{i\frac{\theta}{2}}+ f_2(r)e^{-i\frac{\theta}{2}}\r), \quad \Phi_2(r,\theta) = i\l(f_1(r) e^{i\frac{\theta}{2}} - f_2(r) e^{-i\frac{\theta}{2}}\r).
\end{eqnarray}
  After inserting the expressions  in Eq.~(\ref{Fphitz}) into the effective action in  Eq.~(\ref{eff1}) we find
\begin{eqnarray}
\mathcal{L}_{\rm eff} &&= \Tr F_{i\a} F_i^\a + \Tr \l(D_\a \Phi\r)^\dagger D^\a \Phi\nonumber\\
&&= \l[\l( \p_i \Psi_1 + \zeta  b_i\Psi_2  \r)^2 + \l( \p_i\Psi_2 - \zeta b_i  \Psi_1\r)^2  + g^2\xi^2 \l| \Psi_1 \,\Phi_2   +  \Psi_2 \, \Phi_1 \r|^2\r]
\frac{\p_\a \varphi\p^\a \varphi}{2 g^2}. \quad \label{eff2}
\end{eqnarray}
It can be observed here that the derivative terms of the above effective action is invariant under an $SO(2)$ transformation generated by $\scriptstyle{T =
\left(
\begin{array}{cc}
0  &  1  \\
-1  &  0   
\end{array}
\right)}$. 
This leads us to  define a complex function 
\begin{align}
\Psi(r, \theta) = \Psi_1(r, \theta) + i \Psi_2(r, \theta).
\end{align}
By using this complex function, 
the above effective action in Eq.~(\ref{eff2}) can be rewritten in a simpler form as
\begin{align}\label{Ipsi}
\mathcal{I}_{\rm eff} &=\int d^2x\l[ \l|D_i \Psi\r|^2 + \Delta_g^2 \l| \Psi^* q_1 - \Psi q_2\r|^2\r]\int dt dz\left\{ \frac{\p_\a \varphi\p^\a \varphi}{2 g^2}\r\} ,\nonumber\\ 
 &= I_\Psi \int dt dz\left\{ \frac{\p_\a \varphi\p^\a \varphi}{2 g^2}\r\}, \qquad  \Delta_g^2 = \xi^2 g^2,
 \end{align}
where $D_i\Psi = (\p_i -i \zeta b_i)\Psi$ and
\begin{eqnarray}
I_\Psi = \int d^2x\l[ \l|D_i \Psi\r|^2 + \Dg^2 \l| \Psi^* q_1 - \Psi q_2\r|^2\r],\qquad q_1 = f_1 e^{i\half \theta}, \quad q_2 = f_2 e^{-i\half \theta}.
\end{eqnarray}
We should notice  here that $\Psi$ behaves as if a scalar field in $1 +1$ dimensions in the background of the vortex Abelian gauge field $b_i$ with a charge $\zeta = \half $.  The important is that the $(x, y)$ and $(t, z)$ dependent parts  are  completely separated. So we may integrate over $x$-$y$ plane and get the $1+1$ dimensional vortex effective action.  To evaluate the front factor we extremize  $I_\Psi $  by  varying $\Psi$  and find the equation for $\Psi $ as
\begin{eqnarray}
\label{modeq1}
&&D_i^2 \Psi - \Dg^2\l[\l(f_1^2 + f_2^2\r) \Psi - 2 f_1 f_2 e^{i\theta} \Psi^*\r] = 0 .
\end{eqnarray}
We may write the equation in the polar coordinates $(r , \theta)$ as
\begin{eqnarray}
\label{modeq2}
\frac{1}{\rh}\frac{\p}{\p \rh} \l(\rh \frac{\p \Psi}{\p \rh} \r)+ \l(\frac{\p_\theta - i \zeta A(\rh)}{\rh} \r)^2\Psi - \l[\l(f_1^2 + f_2^2\r) \Psi - 2 f_1 f_2 e^{i\theta} \Psi^*\r] = 0 ,
\end{eqnarray}
where we have made the above equation dimensionless by defining a rescaled length  as $\rh = \Delta_g r$.

In order to find solutions of this equation, let us decompose  the complex profile function $\Psi$ into partial wave modes as
\begin{eqnarray}
\label{siex}
\Psi(\rh, \theta) = \sum_{m\ge0} \l[ a_m(\rh) e^{im \theta} + b_m(\rh) e^{-i m \theta}\r], \qquad a_0(\rh) =0,
\end{eqnarray}
for $m =0$,  $\Psi$ would only be described by $b_0(\rh)$.  After inserting  expansion of Eq.~(\ref{siex}) into Eq.~(\ref{modeq2}) we find the coupled equations 
of the infinite tower of pairs $\{a_m(\rh), b_m(\rh)\}$ as
\begin{eqnarray}
\label{ameq}
&&\frac{1}{\rh}\frac{\p}{\p \rh} \left(\rh \frac{\p a_m}{\p \rh}\right) - \l(\frac{m -  \zeta A(\rh)}{\rh} \r)^2 a_m - \Big[\l(f_1^2 + f_2^2\r) a_m - 2 f_1 f_2 b_{m-1}\Big] = 0,\\
&&\frac{1}{\rh}\frac{\p}{\p \rh} \left(\rh \frac{\p b_m}{\p \rh}\right) - \l(\frac{m + \zeta A(\rh)}{\rh} \r)^2 b_m - \Big[\l(f_1^2 + f_2^2\r) b_m - 2 f_1 f_2 a_{m+1}\Big] = 0.
\label{bmeq}
\end{eqnarray}

From now on, we would like to show the following 
expressions give a solution to the above equations  (\ref{ameq}) and (\ref{bmeq}): 
 \begin{eqnarray}
 \label{ambm}
a_m(\rh)  &=& \rh^{m-\zeta} f(\rh)^{\zeta}, \qquad \text{for}\,\, m > 0 ,\nonumber\\ b_m(\rh) &=&  \rh^{m+\zeta} f(\rh)^{- \zeta},  \qquad  \text{for}\,\, m \ge 0. \label{eq:solution}
\end{eqnarray}
Inserting these solutions into the last terms of Eqs.~(\ref{ameq}) and (\ref{bmeq}) yields 
(by using $1-\zeta = \zeta$)
\begin{eqnarray}
&&\l[\l(f_1^2 + f_2^2\r)a_m - 2 f_1 f_2   b_{m-1}\r]  =    \l[f_1^2 - f_2^2\r]  a_m(\rh), \nonumber\\
&&\Big[\l(f_1^2 + f_2^2\r) b_m - 2 f_1 f_2 a_{m+1}\Big]  = - \l[f_1^2 - f_2^2\r]  b_m(\rh)
\end{eqnarray}
where we have used the identity $f_1 (\rh)   f(\rh)^{- \zeta} = f_2 (\rh)    f(\rh)^{1 - \zeta} =  f_2(\rh)   f(\rh)^{\zeta}$. Then, we have 
\begin{eqnarray}
\label{ameq2}
&&\frac{1}{\rh}\frac{\p}{\p \rh} \left(\rh \frac{\p a_m}{\p \rh}\right) - \l(\frac{m -  \zeta A(\rh)}{\rh} \r)^2 a_m - \l[f_1^2 - f_2^2\r]  a_m(\rh) = 0,\nonumber\\
&&\frac{1}{\rh}\frac{\p}{\p \rh} \left(\rh \frac{\p b_m}{\p \rh}\right) - \l(\frac{m + \zeta A(\rh)}{\rh} \r)^2 b_m + \l[f_1^2 - f_2^2\r]  b_m(\rh)= 0.
\end{eqnarray}
In order to show that these equations hold,  
let us use the BPS equations.  Following Ref.~\cite{Chatterjee:2017jsi} we may express the  BPS equations for the profile functions $f_1(\rh), f_2(\rh)$ and $h_A(\rh) = 1 - A(\rh)$  as 
\begin{align}\label{BPS3}
 \frac{1}{\rh}\frac{\p}{\p \rh} \l(h_A(\rh)\r) - 2 \l[f_1(\rh)^2 - f_2(\rh)^2 \r] =0,\qquad h_A(\rh) = \rh \frac{\p}{\p \rh}\log f(\rh).
\end{align}
where $f(\rho)$ is defined as $f(\rh)= \frac{f_1(\rh)}{f_2(\rh)}$.
By defing a function $f_m(\rh) = {\rh}^{m-\xi} f(\rh)^\xi$, 
this equation can be rewritten as 
\begin{eqnarray}\label{fm1}
\rh \frac{\p}{\p \rh} f_m(\rh) =  \Big(m - \xi A(\rh)\Big) f_m(\rh).
\end{eqnarray}
where $\xi$ is any real number. By applying $\l(\frac{1}{\rh} \p_\rh\r)$ from the left and by using  
Eq.~(\ref{BPS3}) again, this equation can be rewritten as
\begin{eqnarray}
\frac{1}{\rh} \frac{\p}{\p \rh} \Big(\rh \frac{\p}{\p \rh} f_m(\rh)\Big) 
 = \l( \frac{m  - \xi  A(\rh)}{\rh}\r)^2  f_m(\rh) +  2\xi \Big[f_1(\rh)^2 - f_2(\rh)^2 \Big] f_m(\rh). 
\end{eqnarray}
So for $\xi = \pm\zeta$, this reduces to
\begin{eqnarray}\label{fm2}
\frac{1}{\rh} \frac{\p}{\p \rh} \Big(\rh \frac{\p}{\p \rh} f_m(\rh)\Big) -  \l(\frac{m \mp  \zeta  A(\rh)}{\rh}\r)^2  f_m(\rh) \mp 2 \zeta  \l[f_1(\rh)^2 - f_2(\rh)^2 \r] f_m(\rh) =0.
\end{eqnarray}
This equation is the same with Eq.~(\ref{ameq2}) if we identify 
$f_m$ with $a_m$ or $b_m$ for $\zeta = \half$. 
We thus have found an exact solution to Eq.~(\ref{ameq2}) is 
given by Eq.~(\ref{eq:solution}).

Let us discuss the asymptotic behaviors at large and short distances. 
At large distances the solutions diverge as
\begin{eqnarray}
&& a_m(\rh)  \longrightarrow  \rh^{m-\zeta} , \qquad \text{for}\,\, m > 0 ,\nonumber\\ && b_m(\rh) \longrightarrow  \rh^{m+\zeta} ,  \qquad  \text{for}\,\, m \ge 0, 
\end{eqnarray}
since $ f(\rh) \longrightarrow 1$ as ${\rho\rightarrow \infty}$. This can also be understood by analyzing asymptotic behavior of Eq.~(\ref{ameq}).   Since at large distances $f_1(\rh) = f_2(\rh)$, the Eq.~(\ref{ameq}) has the solution $b_m(\rh) = a_{m+1}(\rh) \simeq \rh^{m+\zeta} $. 
This behavior at large distance is known in the literature \cite{Alford:1990ur}.

On the other hand, the short distance behavior was not known before.
To describe  the asymptotic behavior of the solution one should notice that at the center of the vortex the solution vanishes as $\sim \rh^m$ except for $b_0$.  The $\rho$ dependence cancels for $b_0$ in the limit  $\rho \longrightarrow 0$ since $ f_1(\rho)\sim \rho$ as $\rho \longrightarrow 0$ and we find
\begin{eqnarray}
b_0 \longrightarrow f_2(0)^{\half}, \qquad \text{as}\,\,\, \rho \longrightarrow 0,
\end{eqnarray}
where $f_2(0)\ne 0$ can be observed from the solution plot in Fig.~\ref{profile}.


Finally, let us analyze the energy by using the solution we found. 
The energy of the fluctuation mode is determined with the front factor $I_\Psi$ defined in Eq.~(\ref{Ipsi}) as
\begin{eqnarray}
\mathcal{E} = I_\psi \left\{ \frac{(\p_\a \varphi)^2}{2 g^2}\r\}. 
\end{eqnarray}
To derive the value of $I_\Psi$ we have to use the equations of motion of $\Psi$ written in Eq.~(\ref{modeq1}). We multiply $\Psi^*$ with 
Eq.~(\ref{modeq1}) and add this with its complex conjugate equation 
to yield 
\begin{eqnarray}
\Psi^*D_i^2\Psi + \l(D_i^2 \Psi\r)^*\Psi = 2 \Dg^2 \l| \Psi^* q_1 - \Psi q_2\r|^2 .
\end{eqnarray}
This relation would help us to compute $I_\Psi$ exactly as
\begin{eqnarray}
I_\Psi &=&  \int d^2x\l[ \l|D_i \Psi\r|^2 + \Dg^2 \l| \Psi^* q_1 - \Psi q_2\r|^2\r]
= \half  \int d^2x\,\, \nabla^2\l(\Psi^* \Psi\r).
\end{eqnarray}
If we consider the system within a large loop of dimensionless radius $R$, 
the above integral gives 
\begin{eqnarray}
 I_\Psi = \pi R +  \pi \sum_{m> 0}\l[(2m -1)R^{2m-1} + (2m+1) R^{2m+1}\r].
\end{eqnarray}
So the energy of the zero mode can be expressed as 
\begin{eqnarray}
 \mathcal{E} = \frac{2\pi R }{4 g^2}\left[  1 +   \sum_{m> 0}\l\{(2m -1)R^{2(m-1)} + (2m+1) R^{2m}\r\}\r]  (\p_\a \varphi)^2. 
 \end{eqnarray}
 For a zero mode of wavelength $\lambda$ in the $z$-direction the 
 energy ($m =0$) behaves as
 \begin{eqnarray}
 \mathcal{E_\lambda} = \frac{2\pi R }{4 g^2 \lambda^2}.
\end{eqnarray}
So for fixed radius $R$ energy of very long wavelength zero mode fluctuations vanish.

\section{Summary and Discussion}\label{discussion}
In this paper, we have studied 
the zero modes of a single BPS Alice string in the $U(1)_b\times SU(2)$ gauge theory with one charged complex adjoint scalar field. In this system, the $U(1)_b\times SU(2)$ symmetry is spontaneously broken to $ \mathbb{Z}_2\ltimes U(1) \simeq O(2)$ and this symmetry breaking can create 
Alice strings. This Alice string has a peculiar property that the generator of the unbroken $U(1)$ symmetry changes its sign as it encircles the string once, due to the $\mathbb{Z}_2$ factor of a semi-direct product in the unbroken symmetry. 
We constructed this vortex after the BPS completion in the previous paper.
We have discussed the $U(1)$ zero mode as well as the translational modes of a single BPS Alice string. The translational modes are found to be similar to other conventional vortices. The $U(1)$ mode is generated due to the spontaneous breaking of the unbroken $U(1)$ bulk symmetry inside the vortex core. 
By promoting the moduli parameters as fields in the ($t,z$) plane 
and by solving the Gauss law equation,
we have written down the effective action. 
To solve the equation, we have expanded the profile function in partial wave modes and have found that the equations are exactly solvable in terms of vortex profile functions $f_1(r)$ and $f_2(r)$. The partial wave modes have been found to be divergent in powers of radius. After inserting the zero mode profile functions into the effective action, we have written down the effective energy, which is found to be also divergent. The minimum mode is divergent linearly as expected from the derivation in Ref.~\cite{Alford:1990mk}.

In our calculation we have not taken into account the massless bulk gauge field fluctuation effect. 
In principle, to write down an effective action by using renormalization group, one should not integrate it out since it is massless.  
In this situation, by following Ref.~\cite{Witten:1984eb}, one may conjecture to write down a $2$d-$4$d action as (see Ref.~\cite{Bolognesi:2015ida})
\begin{eqnarray}
\mathcal{I}_{2d-4d} = -\int d^4x \frac{1}{4} F_{\mu\nu}^2 + M_r^2\int dtdz \l(\p_\alpha \varphi + A_\alpha \r)^2,
\end{eqnarray}
where $M_r$ is the regularized  mass and the field strength can be defined with a branch cut as
\begin{eqnarray}
F_{\mu\nu} 
= \p_\mu A_\nu -\p_\nu A_\mu - 2\pi \zeta \Sigma_{\mu\nu}.
\end{eqnarray}
Here, $\Sigma_{\mu\nu}$ is a branch cut orthogonal to string world-sheet 
whose direction is determined by a singular gauge and 
can be changed by a singular gauge transformation. 
The electric charge changes its sign when it passes across the cut.  
The coefficeint $2\pi\zeta$ is an Aharonov-Bohm (AB) phase 
and in our case $\zeta= \half$.  
Then, the AB scattering of photons off from an Alice string  and 
AB scattering of two Alice strings should be able to be studied. 
In particular for the latter, it can be dealt within the moduli approximation 
due to the BPS nature.

In this paper, we have discussed bosonic zero modes a bosonic theory. 
There appear fermion zero modes in the string core, 
once we couple the theory to fermions, such as supersymmetric extensions. 
Fermions zero modes will give rise to a non-Abelian statistics
for both Majorana \cite{Ivanov:2000mjr,Yasui:2010yh}  and Dirac fermions \cite{Yasui:2011gk}. It will be novel since there are two origin of non-Abelian statistics: bosons and fermion zero modes.

In this paper we have discussed zero mode excitations of 
a single Alice string and have found that 
the $U(1)$ zero mode is non-normalizable. 
It would be more interesting to study multi-string configurations because for even number of strings there will be no obstruction globally. 
So one may expect to have a normalizable $U(1)$ mode for even number of strings, and that the system may possess finite electric field which is expected to explain the existence of a delocalized Cheshire charge \cite{Alford:1990ur}. 
Since our Alice strings are BPS, 
we can consider a stable multi-vortex system.
A similar zero mode whose wave functions are spread between 
solitons are known for BPS semi-local vortices \cite{Eto:2007yv}.

It is  also important to prove the index theorem for these vortices. 
It would be different from that of conventional vortices in the Abelian-Higgs model, since there is a massless field present in the bulk. 
So the vanishing theorem does not work in this case. We keep these issues as our future problems.

\section*{Acknowledgement}

This work is supported by the Ministry of Education, Culture, Sports, Science (MEXT)-Supported Program for the Strategic Research Foundation at Private Universities ``Topological Science'' (Grant No. S1511006). 
C.~C. acknowledges support as an International Research Fellow of the Japan Society for the Promotion of Science (JSPS)(Grant No: 16F16322).
The work of M.~N. is supported in part by 
JSPS Grant-in-Aid for Scientific Research (KAKENHI Grant No. 16H03984), 
and by a Grant-in-Aid for
Scientific Research on Innovative Areas ``Topological Materials
Science'' (KAKENHI Grant No.~15H05855) 
from the the Ministry of Education,
Culture, Sports, Science (MEXT) of Japan.

\end{document}